\documentstyle[pre,aps,multicol]{revtex}

\draft

\begin{document}

\title{Surface Critical Behavior in Systems with Absorbing States}

\author{Kent B{\ae}kgaard Lauritsen,$^1$ Per Fr\"ojdh,$^{2,3}$ and
        Martin Howard$^1$}
\address{$^1$Niels Bohr Institute, Center for Chaos and Turbulence Studies, 
        Blegdamsvej 17, DK-2100 Copenhagen, Denmark}
\address{$^2$Department of Physics, Stockholm University, Box 6730,
        S-113 85 Stockholm, Sweden}
\address{$^3$NORDITA, Blegdamsvej 17, DK-2100 Copenhagen, Denmark}

\date{\today}

\maketitle

\begin{abstract}
We present a general scaling theory for the surface critical behavior of
non-equilibrium systems with phase transitions into absorbing states.
The theory allows for two independent surface 
exponents which satisfy generalized hyperscaling relations. 
As an application we study a generalized version of directed 
percolation with two absorbing states. We find two distinct surface 
universality classes associated with inactive and reflective walls.
Our results indicate that the exponents associated with
these two surface universality classes are closely connected.
\end{abstract}

\pacs{PACS numbers: 05.40.+j, 64.60.Ht, 68.35.Rh}

\begin{multicols}{2}
\narrowtext


The critical behavior of systems with boundaries has been the 
focus of much research in recent years \cite{review_surf}. 
So far most work on surface critical behavior and
on the analysis of surface universality classes has been within the framework
of equilibrium statistical mechanics. However, the same ideas and principles 
also apply to non-equilibrium systems. 
A prominent example of such a non-equilibrium process is directed percolation
(DP), which is the generic model for systems with a non-equilibrium
phase transition from a state with activity (e.g., with a nonzero
density of particles)
into a so-called absorbing state (with zero activity).
An understanding of DP is important for 
a wide variety of different systems encompassing 
epidemics, chemical reactions, 
interface pinning/depinning,
spatio-temporal intermittency,
the contact process, and certain cellular automata \cite{dickman}.
Recently, however, studies have been made of a number of systems with 
absorbing states which do not belong to the DP class. One prominent example
is a particular reaction-diffusion model called
branching and annihilating random walks with an even number of
offspring (or BAW for short, where in this paper BAW refers exclusively to
the even offspring case) \cite{jensen:1994}.
Other systems in the BAW class (at least in $1+1$ dimensions) 
include certain probabilistic cellular automata
\cite{automata}, monomer-dimer models \cite{monomerdimer},
non-equilibrium kinetic Ising models \cite{menyhard-odor}, and 
generalized DP with two absorbing states (DP2) \cite{haye}.

In this paper we address the impact of walls on systems with phase
transitions into absorbing
states. We have developed a general scaling theory 
which allows for two independent surface exponents, which
satisfy generalized hyperscaling relations. As an application, we have
investigated the surface critical behavior 
of DP2. Our numerics indicate that
DP2 exhibits a far richer surface structure than DP:
we find two different surface universality classes
for DP2 with inactive and reflective walls,
and our numerical results indicate that the exponents
associated with these two classes are closely connected. These results can
be successfully contained within our scaling theory.
However, we emphasize that the theory is much more
general than this and should also apply to other types of 
systems with walls and absorbing states, e.g., to surface effects in catalytic
reactions and systems 
exhibiting self-organized criticality \cite{btw}.

Before turning to the surface critical behavior of DP2 (in $1+1$ dimensions)
and BAW, we
begin by discussing the main features of the corresponding bulk systems 
and then identify some differences and similarities with DP. 
Many models in the BAW class 
\cite{jensen:1994,automata,monomerdimer,menyhard-odor} 
conserve particle number modulo 2, but this appears not to be the
fundamental requirement for the emergence of the new universality class.
Instead the key underlying feature seems to be the presence of a symmetry
relating the various absorbing states \cite{hwang-etal}. This has been 
further demonstrated by Hinrichsen who introduced a generalized
version of the Domany-Kinzel model with $n$ absorbing states \cite{haye}.
This model, which we will refer to as DP$n$, is defined on a 
$d$-dimensional lattice (in space). At time $t$, the state
$s^{t}_{i}$ of the $i$-th site can be 
either active ($A$) or in one of $n$ inactive states ($I_1$, \ldots, $I_n$). 
In $1+1$ dimensions,
the update probabilities $P(s^{t+1}_{i} | s^{t}_{i-1}, s^{t}_{i+1})$
are given by
$P(I_k|I_k,I_k)=1$, $P(A|A,A)=1-nP(I_k|A,A)=q$,
$P(A|I_k,A)=P(A|A,I_k)=p$,
$P(I_k|I_k,A)=P(I_k|A,I_k)=1-p$, $P(A|I_k,I_l)=1$, where
$(k,l=1,\ldots,n;k\neq l)$ 
(see also \cite{haye} for a more complete explanation of the model).
For $n=1$ these rules are equivalent to the 
Domany-Kinzel model which belongs to the DP universality class  
(apart from one special point which belongs
to the compact DP universality class) \cite{domany1,domany2}.
For $n \geq 2$, the distinction between regions of different inactive 
states is preserved by demanding that they are separated by active ones.
Monte Carlo simulations show that bulk DP2 belongs to the bulk
BAW class in $1+1$ 
dimensions \cite{haye}, whereas this probably does not hold in higher 
dimensions.

The growth of both BAW and DP clusters in the bulk close to criticality 
can be summarized by a set
of independent exponents. A natural choice is to consider 
$\nu_\perp$ and $\nu_\parallel$ which describe the divergence of the 
correlation lengths in space, 
$\xi_\perp \sim |\Delta|^{-\nu_\perp}$, and time $\xi_\parallel
\sim |\Delta|^{-\nu_\parallel}$, where $\Delta \equiv p - p_c$
describes the deviation from criticality. 
We also need the order parameter exponent $\beta$, which can 
be defined in two a priori different ways: it is either governed by 
the percolation probability (the probability that a cluster grown from a 
finite seed never dies),
\begin{equation}
\label{P_bulk}
         P(\Delta) \sim \Delta^{\beta_{\rm seed}}, \qquad \Delta > 0,
\end{equation}
or by the density of active sites in the steady state,
\begin{equation}
\label{n(Delta)}
         n(\Delta) \sim \Delta^{\beta_{\rm dens}}, \qquad \Delta > 0.
\end{equation}
For the case of DP, it is known that $\beta$ is unique: 
$\beta_{\rm seed} = \beta_{\rm dens}$ in any dimension. 
This follows from theoretical considerations 
\cite{grassberger-torre,cardy-sugar} and has been verified by extensive 
numerical calculations. The relation also holds for BAW in $1+1$ dimensions,
a result first suggested by numerics and now backed up by an exact
duality mapping \cite{mussawisade-etal}. However, this exponent equality
is certainly not always true---for example it breaks down for certain 
systems with infinitely many absorbing states \cite{mendes,munoz}.

Furthermore, $\beta_{\rm seed} \neq \beta_{\rm dens}$ for BAW in high 
enough dimension: if we consider the mean-field regime valid for
spatial dimensions $d>d_c=2$, then
the system is in an inactive state only for a zero branching rate,
whereas any non-zero branching rate results in an active state. The 
steady-state density (\ref{n(Delta)}) approaches zero continuously 
(as the branching rate is reduced towards zero)
with the mean-field exponent $\beta_{\rm dens}=1$. 
Nevertheless, for $d>2$ the survival probability 
(\ref{P_bulk}) of a particle cluster will be finite for {\it any\/} 
value of the branching rate, implying that 
$\beta_{\rm seed}=0$ in mean-field theory. This result follows from 
the non-recurrence of random walks in $d>2$. 

{}From the perspective of formulating field theories for BAW, the
$1+1$ dimensional case poses considerable difficulties \cite{cardy-tauber}.
These stem from the presence of two critical dimensions: $d_c=2$ (above
which mean-field theory applies) and $d_c' \approx 4/3$ (where for
$d>d_c'$ the branching reaction is a relevant process at the pure
annihilation fixed point, whereas for $d<d_c'$ it is irrelevant there
\cite{cardy-tauber}). This means
that the (physically interesting) spatial dimension $d=1$ cannot be accessed 
using controlled expansions down from the upper critical dimension 
$d_c=2$.
However if we assume that a (bulk) scaling theory can be properly justified 
(as it can be for DP, and BAW for $d>d_c'$), then it is straightforward 
to relate the above set of exponents to those of other quantities.
Keeping the distinction between $\beta_{\rm seed}$ and $\beta_{\rm dens}$, 
the average lifetime of finite clusters, 
$\langle t \rangle \sim |\Delta|^{-\tau}$, satisfies 
$\tau = \nu_\parallel - \beta_{\rm seed}$, and the average mass of finite 
clusters, 
\begin{equation}
        \langle s \rangle \sim |\Delta|^{-\gamma} ,
                        \label{eq:<s>}
\end{equation}
leads to the following hyperscaling relation:
\begin{equation}
        \label{gen_bulk_hyperscaling}
        \nu_{\parallel} +d\nu_{\perp} 
                = \beta_{\rm seed} + \beta_{\rm dens} +\gamma.
\end{equation} 
Note that (\ref{gen_bulk_hyperscaling}) is consistent with the
distinct upper critical dimensions for BAW and DP. Using the above
mean-field values for BAW and $\nu_\perp=1/2$, $\nu_\parallel=1$, and 
$\gamma=1$, we verify $d_c = 2$. 
In contrast, for DP one has the mean-field exponent $\beta_{\rm seed} = 1$ 
and $d_c = 4$.

We now turn to the surface critical behavior of DP2 and
show how the above relations and exponents are modified in a semi-infinite 
geometry where we place a wall at 
$x_\perp=0$ [${\bf x}=({\bf x_{\parallel}},x_{\perp})$, with the $\perp$
and $\parallel$ directions being relative to the wall].
In the simulations we start from an absorbing state,
where all sites are in the state $I_1$. We then initiate a cluster 
by placing a seed (site in state $A$) next to the wall.
However, the analogy with DP is no longer immediate, as our numerical 
measurements in $1+1$ dimensions indicate that DP2 supports an additional 
surface exponent as well as an additional surface universality class. 
The type of surface universality class is governed by the 
choice of boundary condition (BC). We have studied two types of
BC: the inactive BC (IBC) where the wall sites are always
in the inactive state $I_1$, and the reflective BC
(RBC), where the wall acts like a ``mirror'' by letting imaginary
sites next to the outer side of the wall be the mirror images of those
on the inside. 

By growing a DP cluster near an IBC wall, it has been observed numerically
in $d=1,2$ that certain exponents are altered 
\cite{essam-etal:1996,lauritsen-etal}. This behavior has been explained 
by a scaling theory \cite{dp-wall-edge} that explicitly takes surface 
critical phenomena into account and connects IBC with the {\em
ordinary\/} transition \cite{janssen-etal}. 
Apart from the above (three) independent bulk exponents, 
an additional universal surface exponent 
must be included, which satisfies a generalized hyperscaling relation
\cite{dp-wall-edge}. The survival probability (\ref{P_bulk}) for a cluster
started close to the wall has the form
\begin{equation}
        P_1(t,\Delta) 
        = \Delta^{\beta_{1, \rm seed}\,} \psi_1(t/\xi_{\parallel}), 
        \qquad \Delta > 0,
                                \label{P1(Delta)}
\end{equation}
where the subscript `1' refers to the wall. However, in analogy with 
the bulk case, an order parameter can also be defined by
the density of active sites on the wall in the steady state:
\begin{equation}
        n_1(\Delta) \sim \Delta^{\beta_{1, \rm dens}} ,
        \qquad \Delta > 0. 
                                \label{n_1(Delta)}
\end{equation}
More generally the steady-state density (\ref{n(Delta)}) is now given by
   $n(\Delta,x_\perp) = \Delta^{\beta_{\rm dens}}\,
     \varphi(x_\perp/\xi_\perp)$,
where the scaling function $\varphi$ behaves in such a way that
   $n(\Delta,x_\perp)$ for $x_{\perp}/\xi_{\perp} \ll 1$
crosses over to the surface behavior (\protect\ref{n_1(Delta)}).

For the case of DP, the surface exponents fulfill
$\beta_{1, \rm seed} = \beta_{1, \rm dens}$, 
as can be shown by a field-theoretic derivation of an appropriate 
correlation function \cite{dp-wall-edge}. 
However, for DP2 this exponent equality is no longer true.
Our numerical results in $1+1$ dimensions yield two distinct surface
exponents, $\beta_{1, \rm seed} \neq \beta_{1, \rm dens}$, 
although the corresponding bulk exponents coincide, as expected.
The values of these surface exponents depend on the boundary conditions 
and by changing from IBC to RBC or vice versa, we observe that the 
assignment of the exponents is interchanged (see below). 
Further investigations are needed in order to determine whether
the wall may have broken a (duality) symmetry present
in the bulk (which forces the bulk exponents to coincide) and whether
the operation of this symmetry relates IBC to RBC and vice versa.
In contrast for surface DP, we note that
IBC and RBC belong to the same surface universality class.
 
By keeping $\beta_{1, \rm seed}$ and $\beta_{1, \rm dens}$ distinct, 
we can now set up a general scaling theory for the surface critical behavior 
in systems with absorbing states. An ansatz for the coarse-grained density 
of active sites $\rho_1$ at the point (${\bf x}$, $t$) of a cluster grown
from a single seed located next to the wall, has the form
\begin{equation}
        \label{ansatz_rho_wall}
        \rho_{1}(x,t,\Delta) = \Delta^{\beta_{1, \rm seed} + \beta_{\rm dens}}
        f_1  \left(x/{\xi_\perp}, \, {t}/{\xi_\parallel}\right).
\end{equation}
The $\Delta$-prefactor comes from 
(\ref{P1(Delta)}) for the probability that an infinite cluster can be
grown from the seed, and from (\ref{n(Delta)}) for the (conditional)
probability that the point (${\bf x}$, $t$) belongs to this cluster.
The shape of the cluster is governed by the scaling function $f_1$ and
we assume that the density is measured at a finite angle away from the wall.
If the density is measured along the wall, we have instead
\begin{equation}
        \label{ansatz_rho_wall_wall}
        \rho_{11}(x,t,\Delta) =
        \Delta^{\beta_{1, \rm seed} + \beta_{1, \rm dens}}
        f_{11} \left(x/{\xi_\perp}, \,
        t/{\xi_\parallel}\right) ,
\end{equation}
as we pick up a factor $\Delta^{\beta_{1, \rm dens}}$ rather than 
$\Delta^{\beta_{\rm dens}}$ for the probability that (${\bf x}$, $t$) 
at the wall belongs to the infinite cluster. In $1+1$ dimensions,
(\ref{ansatz_rho_wall_wall}) reduces to $\rho_{11}(t,\Delta) = 
\Delta^{\beta_{1, \rm seed} + \beta_{1, \rm dens}} f_{11} (t/\xi_\parallel)$.

Starting from a seed on the wall, the average lifetime of finite clusters,
$\langle t \rangle \sim |\Delta|^{-\tau_1}$, satisfies 
$\tau_1 = \nu_\parallel - \beta_{1, \rm seed}$. The average 
size of finite clusters follows from integrating the cluster density 
(\ref{ansatz_rho_wall}) over space and time:
\begin{equation}
        \label{size_wall}
        \langle s \rangle \sim |\Delta|^{-\gamma_1} ,
\end{equation}
where the surface (susceptibility) exponent $\gamma_1$ is related to
the previously defined exponents via 
\begin{equation}
        \nu_{\parallel}+d\nu_{\perp}=
        \beta_{1, \rm seed}+\beta_{\rm dens}+\gamma_{1} .
        \label{surf_hyperscaling}
\end{equation}
The only difference from (\ref{gen_bulk_hyperscaling}) is that we
have now included a wall. Analogously, by integrating the cluster wall density 
(\ref{ansatz_rho_wall_wall}) over the ($d-1$)-dimensional wall and 
time, we obtain the average (finite) cluster size on the wall,
\begin{equation}
        \label{size_wall_wall}
        \langle s_{\rm wall} \rangle \sim |\Delta|^{-\gamma_{1,1}} ,
\end{equation}
where
\begin{equation}
        \nu_{\parallel}+(d-1)\nu_{\perp}=
          \beta_{1, \rm seed}+\beta_{1, \rm dens}+\gamma_{1,1}.
        \label{surf_hyperscaling_wall_wall}
\end{equation}
Note that if the $\gamma$ susceptibility exponents obtained from 
(\ref{surf_hyperscaling}) and (\ref{surf_hyperscaling_wall_wall})
are negative, then they should be replaced by zero in (\ref{size_wall}) 
and (\ref{size_wall_wall}).
The above scaling theory is generic since it allows for 
$\beta_{1, \rm seed}$ and $\beta_{1, \rm dens}$ to be independent
surface exponents.
When the scaling theory is applied to DP,
it can be fully justified
(with $\beta_{1, \rm seed} = \beta_{1, \rm dens}$) \cite{dp-wall-edge}.
However, if we apply the theory to BAW \cite{baw-wall-mft},
it would again be desirable to obtain a secure renormalization-group
justification for the scaling behavior.
In particular it would be
important to determine from the field-theory whether two
{\em independent\/} surface exponents are present.
However given the fundamental difficulties 
encountered already in the {\it bulk\/} field-theoretic analysis of BAW 
in $1+1$ dimensions \cite{cardy-tauber}, this kind of analysis for the surface
is unlikely to give a complete justification of the scaling theory.

In order to confirm our scaling theory we have performed numerical 
simulations for DP2 in $1+1$ dimensions with walls constrained by IBC or RBC
(see \cite{us} for details). 
We have also performed simulations for DP2 without a wall and obtained 
results for the exponents in complete agreement with \cite{haye}.
There are several estimates for $\beta_{\rm dens}$ ($= \beta_{\rm seed}$) 
available \cite{jensen:1997}: we have used $\beta_{\rm dens}=0.922(5)$ 
\cite{zhong}.

We extract the critical exponents from several measured quantities.
Using (\ref{P1(Delta)}), we find that the survival probability 
for the cluster to be alive at time $t$ has the following behavior at 
criticality ($\Delta = 0$)
\begin{equation}
        P_1(t) \sim t^{-\delta_{1, \rm seed}} ,
        \qquad  \delta_{1, \rm seed} = \beta_{1, \rm seed}/\nu_{\parallel} .
                                          \label{eq:P_1(t)}
\end{equation}
Integrating the densities (\ref{ansatz_rho_wall}) and
(\ref{ansatz_rho_wall_wall}) gives expressions for the
activity at criticality as function of time \cite{us}, e.g.,
\begin{equation}
        N_1(t) \sim t^{\kappa_1},
        \qquad  \kappa_1 = d\chi - \delta_{\rm dens} - \delta_{1, \rm seed}, 
                                        \label{eq:N_1(t)}
\end{equation}
where we have introduced the envelope (or ``roughness'') exponent
$\chi=\nu_\perp/\nu_\parallel$, and 
$\delta_{\rm dens} = \beta_{\rm dens}/\nu_{\parallel}$.
Note that (\ref{eq:N_1(t)}) corresponds to the hyperscaling relation
(\ref{surf_hyperscaling}) at criticality with
$\gamma_1 = \nu_\parallel (1+\kappa_1)$, since
$\langle s \rangle \sim \int dt \, N_{1}(t)$. 
For further confirmations of our numerical data we also considered the
cluster size distributions at criticality. The cluster size $s$ scales
as $s \sim \xi_\perp^d \xi_\parallel n(\Delta) \sim \Delta^{-1/\sigma}$,
with $1/\sigma = d\nu_\perp + \nu_\parallel - \beta_{\rm dens}$.
The probability to have a cluster of size $s$ then reads
\cite{us}
\begin{equation}
   p_1(s) \sim s^{-\mu_1},
   \qquad
   \mu_1 = 1 + \frac{\beta_{1, \rm seed}}
               {d \nu_\perp + \nu_\parallel - \beta_{\rm dens}}.
                                \label{eq:p_1(s)}
\end{equation}

In Table \ref{table-exp} we list our estimates for the critical exponents 
for DP2, where $\delta_{1, \rm dens} = \beta_{1, \rm dens}/\nu_{\parallel}$
is obtained from (\ref{ansatz_rho_wall_wall}) by measuring
the activity at the wall \cite{us}, and $\mu = 1 + \beta_{\rm seed} /
(d \nu_\perp + \nu_\parallel - \beta_{\rm dens})$ corresponds to 
(\ref{eq:p_1(s)}) in the absence of a wall. The results are in
complete accordance with our theoretical analysis: bulk exponents are
unaltered whereas the wall introduces two separate surface exponents.
We have also carried out bulk and surface simulations for $\Delta>0$ and 
confirmed that
our data could be collapsed according to an appropriate survival probability
scaling function [see (\ref{P1(Delta)}) for the surface case], 
using our exponent estimates. This numerically confirms the validity of the 
relation $\delta=\beta/\nu_\parallel$ for the bulk as well as for both sets
of corresponding surface exponents \cite{no-theory-for-it}.
We further observe that the IBC, RBC boundary conditions lead
to {\em different\/} exponents thus showing the existence of two distinct
surface universality classes. Furthermore,
$\beta_{1, \rm seed} \neq \beta_{1, \rm dens}$, although by changing
BCs we observe to good accuracy that 
$\beta_{1, \rm seed}^{(IBC)}=\beta_{1, \rm dens}^{(RBC)}$,
$\beta_{1, \rm seed}^{(RBC)}=\beta_{1, \rm dens}^{(IBC)}$.
As noted above,
this suggests that the two BCs for DP2 are related by a symmetry.
By universality, we expect the same relations to apply to BAW \cite{us}.

By using the explicit definitions of IBC, RBC we can argue that 
$\beta_{1, \rm seed}$ and $\beta_{1, \rm dens}$ should indeed depend on the 
BCs. There will be more activity next to the wall for IBC than for RBC, since
the latter can have regions of $I_2$ located at the wall.
Once created, these $I_2$ regions will survive until the
activity returns to the wall.
Thus, from the wall density (\ref{n_1(Delta)}), it follows 
that $\beta_{1, \rm dens}^{(IBC)} \leq \beta_{1, \rm dens}^{(RBC)}$.
On the other hand, the existence of these $I_2$ regions implies that
the survival probability (\ref{P1(Delta)})
for IBC will be smaller than for RBC, leading 
to $\beta_{1, \rm seed}^{(IBC)} \geq \beta_{1, \rm seed}^{(RBC)}$.
However, from the observation that 
$\beta_{1, \rm seed} + \beta_{1, \rm dens}$ is independent of the BC,
it follows that the average mass on the wall (\ref{size_wall_wall})
is the same for IBC and RBC.
We have also studied several other BCs and found that these give the
same scaling behavior as either RBC or IBC depending on whether or not
the above-mentioned $I_2$ regions can disappear only at the wall or also in
the bulk \cite{us}. 
In terms of the BAW model, however, the distinction between the two BCs is 
slightly different: IBC respects the ``parity'' symmetry of the bulk, 
whereas RBC breaks it. 

For DP it has been customary to investigate
whether the critical exponents can be fitted by (simple) rational
numbers \cite{jensen:1996}. Such a fitting has also been tried for 
bulk BAW with the following guesses in $1+1$ dimensions:
$\kappa = \chi - 2\delta = 0$ and $\chi=4/7$ \cite{jensen:1994}. 
These estimates lead immediately to $\delta=2/7$ (and $\beta/\nu_\perp=1/2$).
It is intriguing to note that our numerical results for DP2
in addition suggest that $\mu_{1} = 3/2$ for IBC and $4/3$ for RBC.
{}From Eq.\ (\ref{eq:p_1(s)}), it then follows that 
$\delta_{1, \rm seed} = 9/14$ for IBC and $3/7$ for RBC.
We would need one more relation in order to obtain the last 
independent exponent. In fact, we observe numerically that 
$
        2\nu_\parallel-\beta_{1, \rm seed}-\beta_{1, \rm dens} = 3 
$
\cite{identity=2},
is valid to within one percent \cite{conjecture}.

In conclusion, we have presented a generic scaling theory of surface
critical behavior in systems with absorbing states. In particular
we have for the first time studied the surface critical behavior
of DP2, a model belonging
to the BAW universality class in $1+1$ dimensions.
Numerical simulations of the DP2 model with two different types 
of boundary conditions have uncovered two surface universality classes. 
Our most important result is that {\it two\/} surface exponents
are required to describe the surface critical behavior.
The results also indicate that the exponents associated with
these two surface universality classes
are closely connected.
We emphasize that our theory is generic for systems with absorbing 
states and therefore should also apply to 
surface effects in, for example,
systems exhibiting self-organized criticality.
It would also be possible to generalize our theory to allow for
edges and corners, which would introduce new exponents and other 
hyperscaling relations.

K.B.L. acknowledges support from the 
Carls\-berg Foundation and P.F. from the Swedish Natural Science
Research Council.

\vspace*{-5mm}
 
\begin{table}[htb]
\caption{Critical exponents obtained from our simulations. 
        For comparison we also list the exponents for DP 
        with an IBC wall in the first column
        \protect\cite{essam-etal:1996,lauritsen-etal,jensen:1996}.
        }
\vspace*{2mm}
\begin{tabular}{||c|c|c|c|c|} \hline
                       &  DP (IBC)   &  DP2      &  DP2 (IBC) &  DP2 (RBC) \\
\hline\hline
$\delta_{\rm dens}$    & 0.159 47(3) & 0.287(5)  &  0.288(2)  & 0.291(4)\\
$\beta_{\rm dens}$     & 0.276 49(4) & 0.922(5)  &  0.93(1)   & 0.94(2) \\
\hline
$\delta_{1,\rm seed}$  & 0.4235(3)   &           &  0.641(2)  & 0.426(3) \\ 
$\beta_{1,\rm seed}$   & 0.7338(1)   &           &  2.06(2)   & 1.37(2)  \\ 
\hline
$\delta_{1, \rm dens}$ & 0.4235(3)   &           &  0.415(3)  & 0.635(2) \\
$\beta_{1, \rm dens}$  & 0.7338(1)   &           &  1.34(2)   & 2.04(2)  \\
\hline
$\mu$                  & 1.108 25(2) &  1.225(5) &            &          \\ 
$\mu_1$                & 1.2875(2)   &           &  1.500(3)  & 1.336(3) \\ 
\hline
\end{tabular}
\vspace*{0.5cm}
\label{table-exp}
\end{table}

\end{multicols}


\vspace{-6mm}

\begin{thebibliography}{99}

\vspace*{-16mm}

\bibitem{review_surf} For reviews see
        K. Binder, in {\it Phase transitions and critical phenomena},
        Vol. 8, ed. by C. Domb and J. Lebowitz (Academic Press,
        London, 1983); H. W. Diehl, in {\it Phase transitions and critical
        phenomena}, Vol. 10, ed. by C. Domb and J. Lebowitz 
        (Academic Press, London, 1986);
        H. W. Diehl, Int. J. Mod. Phys. B {\bf 11}, 3503 (1997).

\bibitem{dickman} 
        R. Dickman, in {\it Nonequilibrium statistical mechanics in one
        dimension}, ed.\ V. Privman 
        (CUP, Cambridge, 1997).

\bibitem{jensen:1994}
        I. Jensen, Phys. Rev. E {\bf 50}, 3623 (1994).

\bibitem{automata} 
        P. Grassberger et al., 
        J. Phys. A {\bf 17}, L105 (1984); 
        P. Grassberger, J. Phys. A {\bf 22}, L1103 (1984).

\bibitem{monomerdimer} 
        H. H. Kim and H. Park, Phys. Rev. Lett. {\bf 73}, 2579 (1994); 
        H. Park et al., Phys. Rev. E {\bf 52}, 5664 (1995).

\bibitem{menyhard-odor}
        N. Menyh\'ard and G. \'Odor, J. Phys. A {\bf 29}, 7739 (1996).

\bibitem{haye} 
        H. Hinrichsen, Phys. Rev. E {\bf 55}, 219 (1997).

\bibitem{btw}
        P. Bak et al., 
        Phys. Rev. Lett. {\bf 59}, 381 (1987); 
        R. Dickman et al.,
        Phys. Rev. E {\bf 57}, 5095 (1998).

\bibitem{hwang-etal}
        W. Hwang et al., 
        Phys. Rev. E {\bf 57}, 6438 (1998).

\bibitem{domany1}
        E. Domany and W. Kinzel, Phys. Rev. Lett. {\bf 53}, 311 (1984).

\bibitem{domany2}
        W. Kinzel, Z. Phys. B {\bf 58}, 229 (1985).

\bibitem{grassberger-torre}
        P. Grassberger and A. de la Torre, 
        Ann. Phys. (N.Y.) {\bf 122}, 373 (1979).

\bibitem{cardy-sugar}
        J. L. Cardy and R. L. Sugar, J. Phys. A {\bf 13}, L423 (1980).

\bibitem{mussawisade-etal} K. Mussawisade et al., J. Phys. A {\bf 31}, 4381 
        (1998).

\bibitem{mendes} 
        J. F. F. Mendes et al.
        J. Phys. A {\bf 27}, 3019 (1994).

\bibitem{munoz}
        M. A. Mu\~noz et al., 
        Phys. Rev. E {\bf 56}, 5101 (1997).

\bibitem{cardy-tauber} 
        J. Cardy and U. C. T\"auber, Phys. Rev. Lett. {\bf 77}, 4780 (1996);
        J. Stat. Phys. {\bf 90}, 1 (1998).

\bibitem{essam-etal:1996}
        J. W. Essam et al.,
        J. Phys. A {\bf 29}, 1619 (1996).

\bibitem{lauritsen-etal}
        K. B. Lauritsen et al.,
        Physica A {\bf 247}, 1 (1997).

\bibitem{dp-wall-edge}
        P. Fr\"ojdh et al., 
        J. Phys. A {\bf 31}, 2311 (1998).

\bibitem{janssen-etal}
        H. K. Janssen et al., 
        Z. Phys. B {\bf 72}, 111 (1988).

\bibitem{baw-wall-mft}
        In BAW mean-field theory:
        $\beta_{1, \rm seed}^{(IBC)}=0$, $\beta_{1, \rm dens}^{(IBC)}=3/2$,
        whereas in DP mean-field theory:
        $\beta_{1, \rm seed}=\beta_{1, \rm dens}=3/2$. 

\bibitem{us}
        P. Fr\"ojdh, M. Howard, and K. B. Lauritsen, unpublished.

\bibitem{jensen:1997}
        I. Jensen, J. Phys. A {\bf 30}, 8471 (1997).

\bibitem{zhong} 
        D. Zhong and D. Ben-Avraham, Phys. Lett. A {\bf 209}, 333 (1995).

\bibitem{jensen:1996}
        I. Jensen, J. Phys. A {\bf 29}, 7013 (1996).

\bibitem{no-theory-for-it}
        Technical difficulties have so far prevented 
        a field-theoretic derivation of this kind of relation for BAW 
        \protect\cite{cardy-tauber}.

\bibitem{identity=2}
        The relation 
        $\nu_\parallel-\beta_{1, \rm seed}=1$ ($\pm 0.0003$) found
        numerically for DP with a wall
        \protect\cite{essam-etal:1996} can be rewritten
        $
                2\nu_\parallel-\beta_{1, \rm seed}-\beta_{1, \rm dens}=2, 
        $
        using the DP relation $\beta_{1, \rm seed}=
        \beta_{1, \rm dens}$. 

\bibitem{conjecture} 
        The other DP2 exponents then follow:
        $\beta_{\rm dens} = \beta_{\rm seed} = 12/13$,
        $\nu_\parallel=42/13$, $\nu_\perp=24/13$; also 
        $\beta_{1, \rm seed}=27/13$ and $\beta_{1, \rm dens}=18/13$ for IBC
        and vice versa for RBC.

\end{thebibliography}
\end{document}